\documentclass[aps,prd,prabib,twocolumn,showpacs,nofootinbib]{revtex4}
\usepackage{graphicx} \usepackage{amsmath} \usepackage{amssymb}
\usepackage{amsfonts} \usepackage{bm}

\begin{document}

\newcommand{\be}{\begin{equation}} \newcommand{\ee}{\end{equation}}
\newcommand{\bea}{\begin{eqnarray}}\newcommand{\eea}{\end{eqnarray}}

\title{Conformal anomaly in non-hermitian quantum mechanics}

\author{Pulak Ranjan Giri} \email{pulakranjan.giri@saha.ac.in}

\affiliation{Theory Division, Saha Institute of Nuclear Physics,
1/AF Bidhannagar, Calcutta 700064, India}

\begin{abstract}
A model of an electron and a Dirac monopole interacting through an
axially symmetric non-hermitian  but $\mathcal{PT}$-symmetric
potential is discussed in detail. The intriguing localization of the
wave-packet as a result of the anomalous breaking of the scale
symmetry  is shown to provide a scale for the system. The symmetry
algebra for the system, which is the conformal algebra $SO(2,1)$, is
discussed and is shown to belong to the enveloping algebra of the
combined algebra, composed of the Virosoro algebra, $\{L_n, n\in
\mathbb{N}\}$ and an abelian algebra, $\{P_n,n\in \mathbb{N}\}$.
\end{abstract}



\date{\today}

\maketitle

\section{INTRODUCTION}
The massless scaler field theory \cite{alfaro} in $(d+1)$-dimensions
with interaction $\mathcal L_{int}= -\mbox{g}\phi^{2(d+1)/(d-1)}$ is
known to have conformal symmetry $SO(2,1)$, generated by dilation
$D$, the Hamiltonian itself $H$ and generator for conformal
transformation $K$. This model defined by the Lagrangian $\mathcal
L=\frac{1}{2}\partial_\mu\phi\partial^\mu\phi-
\mbox{g}\phi^{2(d+1)/(d-1)}$ plays a crucial role in the context of
conformal symmetry in quantum mechanics, when the scalar field is
considered in $(0+1)$-dimension.

Since then a huge number of works \cite {camb,kumar1,giri1,giri3} 
have been reported in
the quantum mechanical settings,  studying   conformal symmetry
\cite{camb1} and related issue like anomaly \cite{giri2}. The basic
ingredient in almost all the cases is the interaction potential of
the form $V_I= Cr^{-2}$ \cite{alfaro,camb,camb1,giri2,feher}. 
The reason for taking the
potential $V_I$ can be understood from the scale transformation
property, $D: t\to \alpha^2 t, D:\boldsymbol{r}\to
\alpha\boldsymbol{r}$, of the potential compared to the kinetic
energy term. The same scale transformation for the potential
$D:V_I\to\alpha^{-2}V_I$ as the kinetic term
$D:\boldsymbol{p}^2/2m\to \alpha^{-2}\boldsymbol{p}^2/2m$ makes the
Lagrangian scales as $D:L\to \alpha^{-2}L$, which is sufficient to
keep the action, $S= \int dtL$, invariant, $D:S\to S$. The
invariance of the system under scale transformation, $D$, has a
consequence on the observables like bound state eigenvalue and
phase-shift of the scattering states. Scale symmetry implies that,
the ground state of the system is not bounded from below, i.e.,
$E_{\mbox{g.s}}=-\infty$. Then the  system is not stable and
therefore will collapse into the singularity. It is however possible
to make these  systems stable against collapse by suitable
quantization. The quantization procedure provide a scale for the
system and shows up as a lower bound to the bound state eigenvalue.

It can be noted that the scale transformation in spherical
co-ordinates, $D:t\to \alpha^2t$, $D:r\to \alpha r$, $D:\theta\to
\theta$ $D:\phi\to\phi$, does not effect the angular coordinates
$\theta$ and $\phi$. One can therefore generalize potential, still
remaining scale covariant, like  $V_{\theta,\phi}=
C(\theta,\phi)r^{-2}$, where now instead of being constat
coefficient,  $C(\theta,\phi)$ is both function of $\theta$ and
$\phi$. Note the scale transformation $D:V_{\theta,\phi}\to
\alpha^{-2}V_{\theta,\phi}$, which is same as the previous scale
covariant potential $V_I$. One can also generalize the kinetic term
to include magnetic vector potential as long as it remains scale
covariant. One can  easily find a magnetic vector potential
$\boldsymbol{A}$, such that the generalized kinetic term
$(\boldsymbol{p}+ e\boldsymbol{A})^2/2m$ transform the same way as
$\boldsymbol{p}^2/2m$. In our present article, we discuss such a
system, where  an electrically charged particle is moving   in the
background field of a magnetic monopole. We also include an
interaction potential, which is axially symmetric
$V_{\mathcal{P}\mathcal{T}}=\frac{c_1}{r(r+z)}+ \frac{c_2}{r(r-z)}$,
where $c_1$, and $c_2$ are two complex valued constant parameters
such that $c_1= c_2^*$. This system is obtained from the generalized
MIC-Kepler system \cite{mar,mar1}, 
which is the system of two dyons with the axially
symmetric potential $V_{MIC}= \frac{c_1}{r(r+z)}+
\frac{c_2}{r(r-z)}- \alpha_s/r + s^2/r^2$.  We  set the Coulomb term
and the extra inverse square term zero, i.e., $\alpha_s= s^2=0$ and
and generalize  the two constants $c_1$ and $c_2$ to complex numbers
in $V_{MIC}$. Note that the complex potential,
$V_{\mathcal{P}\mathcal{T}}$, although makes the system
non-hermitian, it still remains $\mathcal{P}\mathcal{T}$-symmetric.

This article is organized in the following way: We discuss the model
in the next section and offer a physically realizable solution for
the problem. The scale symmetry of the classical version of the
problem is discussed in Sec. III, and  it is shown that scale
symmetry goes anomalous breaking in our quantization process. The
algebraic property of the model is discussed in Sec. IV, where it is
shown that $SO(2,1)$ algebra is a subalgebra of an enveloping
algebra. Finally we conclude in Sec. V.

\section{Electron and Dirac monopole system in $\mathcal{PT}$
-symmetric potential}
The formal Hamiltonian for an electron moving in the background
field of a Dirac monopole and interacting with the potential
$V_{\mathcal{P}\mathcal{T}}$ is written in the form ($\hbar= e= c=
2\times\mbox{reduced mass}=1$)
\begin{eqnarray}
H= (-i\boldsymbol{\nabla} -s\boldsymbol{A})^2 + \frac{c_1}{r(r+z)}+
\frac{c_2}{r(r-z)} \,, \label{electron}
\end{eqnarray}
where according to Dirac quantization condition $s=0, \pm 1/2, \pm
1, \pm 3/2,...$ The vector potential, $\boldsymbol{A}$, due to the
magnetic monopole field, $\boldsymbol{B}= \boldsymbol{r}/r^{-3}$,
has been taken as $\boldsymbol{A}= \left(r^2-rz\right)^{-1}(y, -x,
0)$. Note that we drop the inverse square term
$s^2/(\sqrt{2}r)^{-2}$ from our model Hamiltonian, which was put in
by hand in order to restore the $SO(4)$ and $SO(1,3)$ symmetry for
the bound state and scattering sate respectively.  See Ref.
\cite{zwa} for detail discussion on it. The Hamiltonian
(\ref{electron}), defined on the Hilbert space
$L^2(\boldsymbol{R}^3, r^2drd\Omega)\in \mathcal{H}$ can be
separated in radial and angular part as
\begin{eqnarray}
H\equiv H(r) \oplus r^{-2}\Sigma(\theta,\phi)\,.
\label{Hamiltonian1}
\end{eqnarray}
Note that the radial and angular Hamiltonians  $H(r)=
-\left[r^{-2}\partial_r\left(r^2\partial_r\right)- s^2/2r^2\right]$
and
$\Sigma(\theta,\phi)$  act over the Hilbert spaces
$L^2(\boldsymbol{R}^+, r^2dr)$  and $L^2(S^2,\Omega)$ respectively,
where $L^2(\boldsymbol{R}^3, r^2drd\Omega)\equiv
L^2(\boldsymbol{R}^+, r^2dr)\otimes L^2(S^2,d\Omega)$. We now
consider a similarity transformation (unitary)
$U(r):L^2(\boldsymbol{R}^+, r^2dr)\otimes L^2(S^2,d\Omega)\to
L^2(\boldsymbol{R}^+, dr)\otimes L^2(S^2,d\Omega)$, so that the
radial Hamiltonian is obtained in a convenient form
\begin{eqnarray}
H_U\equiv U(r)^\dagger HU(r)=-\partial_r^2 +(\alpha-s^2)/r^{2}\,,
\label{hamiltonian2}
\end{eqnarray}
where $\Sigma(\theta,\phi)$ has been replaced by its corresponding
eigenvalue $\alpha$, obtained from
$\Sigma(\theta,\phi)Y(\theta,\phi)=\alpha Y(\theta,\phi)$. The
explicit form of the angular Hamiltonian can be found in
\cite{mar,mar1,giri}, but for our present purpose it is not
required. Note that the Hamiltonian (\ref{hamiltonian2}) is a well
known operator appeared in diverse fields in theoretical physics. It
is an example of a class of operators where both the method of
self-adjoint extensions (SAE) and re-normalization technique are
successfully applied in order to get physically realizable
solutions. The specific technique  used depends on the value of the
effective coupling constant $(\alpha-s^2)$ of the inverse square
interaction. In our case $(\alpha-s^2)$ is always is positive. Since
the re-normalization technique is useful for coupling  $<-1/4$
\cite{kumar1}, we rule out the the re-normalization technique from
our consideration because effective coupling is $(\alpha-s^2)>0$.

For potential of the form $V_I= C r^{-2}$ in 1-dimension, one can
show that there is a window in the coupling constant, $-1/4\leq C <
3/4$, where the problem under consideration is not self-adjoint for
a very simple domain and needs a self-adjoint extensions (SAE). Our
model Hamiltonian $H_U$ therefore deserves SAE for $-1/4\leq
(\alpha-s^2)<3/4$. The usual prescription is to define the
Hamiltonian $H_U$ over a very restricted domain
\begin{equation}
\mathcal{D}(H_U)=\{\psi(r)\in
L^2(\boldsymbol{R}^+,dr),\psi(0)=\psi'(0)=0\}\,, \label{domain}
\end{equation}
so that the Hamiltonian $H_U$ easily becomes symmetric,
$(\chi_1,H_U\chi_2)= (H_U\chi_1,\chi_2)$ for $\forall~
\chi_1,\chi_2\in \mathcal{D}_{U}$. Then one needs to  go for a
consistent method to get a SAE for the Hamiltonian $H_U$. We use the
von Neumann's method of SAE for our purpose. It helps us to
construct a self-adjoint domain
\begin{eqnarray}
\mathcal{D}^\omega(H_U)= \{\mathcal{D}(H_U) +
\psi^\omega|\psi^\omega\in \mathcal{D}(H_U^\dagger)\}\,,
\label{domain1}
\end{eqnarray}
where the explicit form of the function $\psi^\omega$ is the linear
combination $\psi^\omega= \psi^+ +\exp(i\omega)\psi^-$ of the two
deficiency space solutions $H_U^\dagger\psi^\pm=\pm i\psi^\pm$
($H_U^\dagger$ is the adjoint of $H_U$). The Hamiltonian $H_U$ is
now self-adjoint over the domain $\mathcal{D}^\omega(H_U)$.

The bound state energy and bound state eigen-function, for $0<
\zeta^2-1/4=\alpha -s^2\leq 3/4$, are respectively given by
\cite{giri3}
\begin{eqnarray}
\nonumber E(L^{-2},\omega)= -L^{-2}\mathcal{F}(\omega)\,,\\
\psi(r)\equiv
K_\zeta\left(\sqrt{|E(L^{-2},\omega)|}r\right)\,,\label{state}
\end{eqnarray}
where $K_\zeta$ is the modified bessel function, $L$ is the length
scale which comes from self-adjoint extensions and
$\mathcal{F}(\omega)$ is a periodic function whose explicit form can
be found by matching the limiting value of the eigenfunction
(\ref{state}) with the domain $\mathcal{D}^\omega(H_U)$ at $r\to 0$,
\begin{eqnarray}
\mathcal{F}(\omega)=\sqrt[\zeta]{\frac{\cos\frac{1}{4}\left(2\omega+
\zeta\pi\right)} {\cos\frac{1}{4}\left(2\omega- \zeta\pi\right)}}\,.
\label{period}
\end{eqnarray}
Note that the periodic function
$\mathcal{F}(\omega)=\mathcal{F}(\omega+\pi)$ also depends on the
coupling constant $\zeta$, besides the SAE parameter $\omega$. The
bound state does not exist for two extremes for the periodic
function, when $|\mathcal{F}(\omega=(1-\zeta/2)\pi)|=0$ (this is the
condition for threshold) or
$|\mathcal{F}(\omega=(1+\zeta/2)\pi)|=\infty$ (this is the condition
when the bound state collapses into singularity).

\section{Anomalous SYMMETRY BREAKING}
We now discuss the scaling symmetry breaking in our model. We start
with the corresponding classical Hamiltonian
\begin{eqnarray}
H_{Cl}= \boldsymbol{\mathcal{D}}_{Cl}^2 +  V_A\,,
\label{ClHamiltonian}
\end{eqnarray}
where now  $\boldsymbol{\mathcal{D}}_{Cl}= (\boldsymbol{p}
-s\boldsymbol{A})$. The lagrangian obtained from the Hamiltonian
(\ref{ClHamiltonian}) is found to be $L_{Cl}= 1/2\boldsymbol{v}^2
-\boldsymbol{A}.{\boldsymbol v}- V_A$. It can be noted that in terms
of dimensions the relation $[H_{Cl}]= [L_{Cl}]= [t^{-1}]=
[\boldsymbol{r}^{-2}]$ evidently makes the action $S=\int L_{Cl}dt$
dimensionless. Consider the scale transformation
$T:\boldsymbol{r}\to \varrho\boldsymbol{r}$, $t\to \varrho^2 t$. The
action $S$ is invariant under this transformation, $T:S\to S$, which
in tern imples the existence of a conserved charge according to the
Noether theorem, known as Dilation  \cite{leblond,cabo}
\begin{eqnarray}
D_{Cl}= \sum\frac{\partial L_{Cl}}{\partial\dot{x}}\Delta x-
T^{00}\Delta t\equiv H_{Cl}t-(1/4)\left[r,p_r\right]_+
\end{eqnarray}
(in symmetrized form), where $T^{00}= \sum\frac{\partial
L_{Cl}}{\partial\dot{x}}\dot{x}-L_{Cl}$. In classical physics the
transformation related to dilation $D_{Cl}$, can be shown
\cite{gozzi,gozzi1} to be  responsible for generating infinitesimal
scale transformation.

In order to see whether the scale symmetry, we just discussed, goes
through unbroken even after quantization of the classical system
$H_{Cl}$, we have to know the possible consequence of the scale
symmetry which we could be able to identify in quantum system. It
can be easily shown that in order the scale symmetry to be unbroken
even after quantization, the system does not have any lower bound of
the energy, which implies that there is no bound state for the
system. The proof goes as follows: Consider the eigenvalue equation
$H\psi(\boldsymbol r)= E\psi(\boldsymbol r)$. The function
$\psi(\varrho\boldsymbol r)$ is also an eigen-state  with eigenvalue
$E/\varrho^2$. This shows that the eigen-state $\psi(\boldsymbol r)$
can be continuously squished towards the center to collapse  to the
singularity in the limit $\varrho\to 0$; $\lim_{\varrho\to
0}\psi(\varrho\boldsymbol r)$, $\lim_{\varrho\to 0}
E/\varrho^2=\infty$.

In our case we showed in the previous section that the model has
single bound state with energy  $E(L^{-2},\omega)$, parameterized by
 $\omega$. Every value of the parameter $\omega$ corresponds to a
well defined boundary condition. The existence of a bound state
indicates that the scale symmetry is anomalously broken
\cite{govinda}. We pointed out in the previous section that there
are two extremes: one is threshold at $(1-\zeta/2)\pi$ and other is
at $(1+\zeta/2)\pi$ where the bound state collapses, indicating that
scaling symmetry still survives  in case of  two inequivalent
quantizations.

\section{The enveloping algebra and its corresponding property}
The model we are discussing in this article has a radial eigen-value
equation (\ref{hamiltonian2}) which possesses $SO(2,1)$ symmetry
generated by the Hamiltonian $H_U$, the dilation $\mathcal{D}= iD$
and the conformal generator $K$. The explicit forms of  two of the
$SO(2,1)$ generators, $H_U$ and $\mathcal{D}$, are known in our case
so far. The explicit form of the generator  $K$ of the algebra
($\hbar=1$) \cite{alfaro}
\begin{eqnarray}
[\mathcal{D},H_U]= H_U, [\mathcal{D},K]= -K, [H_U,K]=
2\mathcal{D}\,, \label{so(2,1)}
\end{eqnarray}
is found to be $K= Ht^2-(1/2)\left[r,p_r\right]_+ + (1/4)r^2$. It
has been shown in the literature \cite{danny} that the Hamiltonian
of form $H_U$ are part of the enveloping  algebra of an algebra
$\mathcal A$, made up with the two sub-algebras. One is the Virasoro
algebra (with generators $L_m, m\in \mathbb{Z}$)
\begin{eqnarray}
\left[L_m,L_n\right]= (m-n)L_{m+n}\,,
 \label{virasoro}
\end{eqnarray}
and other is an abelian algebra (with generators $P_m, m\in
\mathbb{Z}$)
\begin{eqnarray}
\left[P_m,P_n\right]=0\,,
 \label{abelian}
\end{eqnarray}
with  commutators between the  elements of the two different
sub-algebras is
\begin{eqnarray}
\left[L_m,P_n\right]=nP_{n-m}\,.
 \label{combine}
\end{eqnarray}
Consider the representations $L_m= -r^{m+1}\partial_r$ and
$P_m=1/r^m$.  One can write down the above discussed generators of
the $SO(2,1)$ algebra, $H_U$, $\mathcal{D}$ and $K$ in terms of
$L_m$ and $P_n$ for some values of $m$ and $n$,
\begin{eqnarray}
H_U &=&(-L_{-1}+\vartheta P_1)(L_{-1}+ \vartheta P_1)\,,\\
-i\mathcal{D} &=& H_Ut +1/4(L_0+L_{-1}P_{-1})\,,\\ K &=& H_Ut^2
+1/2(L_0+L_{-1}P_{-1})t+(1/2)P_{-2}\,,
 \label{representation1}
\end{eqnarray}
where $\vartheta=\zeta+1/2$. It can be noted that the $SO(2,1)$
generators are products of the elements of the algebra $\mathcal A$
given in Eqs. (\ref{virasoro}), (\ref{abelian}) and (\ref{combine}).
So it does not belong to the  algebra $\mathcal A$, however they
belong to its enveloping algebra $\widetilde{\mathcal A}$. One can
think $SO(2,1)$ as a sub-algebra of the enveloping algebra
$\widetilde{\mathcal A}$. The commutation relation of the
Hamiltonian with the generators of the algebra $\mathcal A$ can be
written as \cite{danny}
\begin{eqnarray}
\nonumber \left[L_n,H_U\right]&=&-
(n+1)\left[L_{n-1},L_{-1}\right]_+ + 2\vartheta(\vartheta-1)
P_{2-n}\,,\\ \left[P_n,H_U\right]&=& n\left[L_{-1},P_{1+n}\right]_+
\label{com}
\end{eqnarray}
The commutation relation of the remaining  $SO(2,1)$ generators with
the elements of the algebra, $\mathcal A$, can be similarly
evaluated as
\begin{eqnarray}
\nonumber \left[L_n,K\right]= \left[L_n,H_U\right]t^2+
 n/2L_n+~~~~~~~~~~~~~~~~~~~
 \\(n+1)/2L_{n-1}P_{-1}-1/2L_{-1}P_{-1-n}-mP_{-2-n}\,,
 \end{eqnarray}
 \begin{eqnarray}
 \left[P_n,K\right]=\left[P_n,H_U\right]t^2-n/2P_n-1/2P_{n+1}P_{-1}~~~\,,
\end{eqnarray}
\begin{eqnarray}
\nonumber \left[L_n,-i\mathcal D\right]=\left[L_n,H_U\right]t+n/42L_n+~~~~~~~~~~~~~~~~\\
(n+1)/4L_{n-1}P_{-1}-1/4L_{-1}P_{-1-n}\,,
\end{eqnarray}
\begin{eqnarray}
\left[P_n,-i\mathcal
D\right]=\left[P_n,H_U\right]t-n/4P_n-1/4P_{n+1}P_{-1}\,.~~
\label{com}
\end{eqnarray}
Note that all the above three commutators are written in terms of
the nonlinear sums of the of elements of $\mathcal{A}$.

\section{CONCLUSION}
In conclusion, we discussed the dynamics of an electron in the field
of a Dirac monopole and interacting with an  axially symmetric
$\mathcal{PT}$-symmetric potential $V_{\mathcal{PT}}$. We find bound state
solutions due to the anomalous breaking of the scaling symmetry of the system
by self-adjoint extensions. We show that the $so(2,1)$ algebra of the system 
belong to the enveloping algebra, $\widetilde{\mathcal A}$, of an algebra, 
$\mathcal A$,  which is a combination of the
Virosoro algebra, $\{L_n, n\in
\mathbb{N}\}$ and an abelian algebra, $\{P_n,n\in \mathbb{N}\}$.

\section{Acknowledgment}
The author thanks P. B. Pal  for his comments on the manuscript.

\end{document}